\documentstyle[12pt,epsfig]{article}
\def\be{\begin{equation}} \def\ee{\end{equation}} \def\bea{\begin{eqnarray}}
\def\eea{\end{eqnarray}} \def\nnb{\nonumber}

\begin{document}

\hfill{May 21, 2022} 

\begin{center}
\vskip 6mm 
\noindent
{\Large\bf  
Elastic $\alpha$-$^{12}$C scattering at low energies 
with the resonant $2_2^+$ and $2_3^+$ states of $^{16}$O
}
\vskip 6mm 

\noindent
{\large 
Shung-Ichi Ando\footnote{mailto:sando@sunmoon.ac.kr}, 
\vskip 6mm
\noindent
{\it
Department of Display and Semiconductor Engineering, \\
Research Center for Nano-Bio Science, \\
Sunmoon University,
Asan, Chungnam 31460,
Republic of Korea
}
}
\end{center}

\vskip 6mm

The elastic $\alpha$-$^{12}$C scattering for $l=2$ at low energies  
is studied in an effective Lagrangian approach. 
We explicitly include two resonant $2_2^+$ and $2_3^+$ states 
of $^{16}$O in the scattering amplitudes and construct an $S$ matrix
assuming that three amplitudes, 
non-resonant part of the amplitude which has the sub-threshold 
$2_1^+$ state of $^{16}$O and those of the two resonant states, 
are represented by the summation of corresponding parts of the phase shift. 
Then, we fit the parameters in the $S$ matrix 
to the phase shift data 
by imposing three conditions at the very low energies where the 
phase shift data are not available.
By using the fitted parameters,
we calculate the asymptotic normalization coefficients (ANC) 
of the $2_1^+$ state of $^{16}$O and 
find that the previously reported small and large values 
of the ANC can be reproduced depending on the imposed conditions, 
but 
we obtain large error bars for the large ANC values which are
reported from the $\alpha$ transfer reactions.  

\vskip 5mm 
\noindent PACS(s): 
11.10.Ef, 
24.10.-i, 
25.55.-e, 
26.20.Fj  

\newpage
\vskip 2mm \noindent
{\bf 1. Introduction}

Radiative $\alpha$ capture on carbon-12, 
$^{12}$C($\alpha$,$\gamma$)$^{16}$O, is an essential reaction in 
nuclear astrophysics, which determines the $^{12}$C/$^{16}$O ratio 
in stars~\cite{f-rmp84}. 
Over the last half-century, many experimental and theoretical studies 
for the reaction have been carried out.
See, e.g., 
Refs.~\cite{bb-npa06,chk-epja15,bk-ppnp16,detal-rmp17,hkvk-rmp20,a-epja21} 
for review. 

Direct measurement of the reaction at the Gamow-peak energy,
$E_G=0.3$~MeV, in stars, 
where $E$ is the kinetic energy of $\alpha$-$^{12}$C system
in the center of mass frame, 
is not easy because the Gamow penetration factor becomes vanishingly small; 
one needs to extrapolate the reaction rate to
$E_G$ by employing a theoretical formula and the experimental data
measured at a few MeV or larger. 
The radiative capture reaction at $E_G$ is known to be $E1$ and $E2$ 
transitions dominant because of the sub-threshold $1_1^-$ and $2_1^+$ 
($J_{i-th}^\pi$) states of $^{16}$O, whose binding energies are
$E(1_1^-)=-0.045$~MeV and $E(2_1^+)=-0.245$~MeV, respectively, from the 
$\alpha$-$^{12}$C breakup threshold energy.   
In our previous works, we have studied elastic $\alpha$-$^{12}$C 
scattering with and without sub-threshold states of $^{16}$O 
for $l=0,1,2,3$~\cite{a-prc18,a-epja16} 
and the inclusion of ground $0_1^+$ state and resonant $0_3^+$ state 
of $^{16}$O for $l=0$~\cite{a-jkps18,a-prc20} along with
the $E1$ transition of $^{12}$C($\alpha$,$\gamma$)$^{16}$O and 
$\beta$ delayed $\alpha$ emission from $^{16}$N~\cite{a-epja21,a-prc19} 
in effective field theory (EFT).
In this work, we study the inclusion of resonant $2_2^+$ and $2_3^+$ states
of $^{16}$O in the calculation of elastic $\alpha$-$^{12}$C scattering
for $l=2$.\footnote{
A preliminary result of this work was reported in Ref.~\cite{a-fbs21}.
} 

A problem of the calculation of elastic $\alpha$-$^{12}$C scattering 
for $l=2$ at low energies in EFT~\cite{a-prc18,a-epja16,klh-jpg13}, 
and also in the other calculations in which 
the effective range expansion is adopted~\cite{oin-prc16}, 
is small values of 
asymptotic normalization coefficient (ANC) of the $2_1^+$ state of $^{16}$O. 
A typical value of the ANC using the effective range
expansion is $|C_b|\simeq 2 \times 10^4$ fm$^{-1/2}$. 
It is significantly smaller than those obtained 
from the $\alpha$-transfer reactions,
$|C_b|=(1.14 \sim 1.82)\times 10^5$ 
fm$^{-1/2}$~\cite{bgkv-prl99,betal-npa07,ab-plb09,aetal-prl15}, 
and those from other theoretical calculations; 
e.g., Sparenberg obtained $|C_b|=1.445\times 10^5$ fm$^{-1/2}$ 
from phase-equivalent super-symmetric potentials~\cite{s-prc04}, 
and Dufour and Descouvemont did $|C_b|=1.26\times 10^5$ fm$^{-1/2}$ from 
generator coordinate method~\cite{dd-prc08}. 
The difference in the ANC values will be consequential 
in an estimate of the $E2$ transition of 
$^{12}$C($\alpha$,$\gamma$)$^{16}$O at $E_G$. 
The ANC values of the $2_1^+$ state of $^{16}$O for other approaches
are well summarized 
in Table XIII in Ref.~\cite{detal-rmp17}
and Table VI in Ref.~\cite{dd-prc08}. 

In this work, we investigate the elastic $\alpha$-$^{12}$C scattering
for $l=2$ at low energies including the resonant $2_2^+$ and $2_3^+$ states
of $^{16}$O in the study so that we can use the whole phase shift data
for $l=2$ at $2.6 \le E_\alpha \le 6.62$~MeV 
reported in Ref.~\cite{tetal-prc09} 
for the parameter fit, where
$E_\alpha$ is the $\alpha$ energy in the lab frame. 
We separate the phase shift into three parts: 
those for the non-resonant part including the sub-threshold $2_1^+$ state 
and for the two resonant parts of the $2_2^+$ and $2_3^+$ states of $^{16}$O.
We also study the inclusion of a contribution 
from the resonant $2_4^+$ state of $^{16}$O
as a background contribution from high energy, to fit a tail of the phase
shift data at the high energy side. 
An aim of the present work is how we can reproduce 
the large ANC values reported in the $\alpha$ transfer reactions 
by using the phase shift data.
For this aim, we introduce three conditions (I), (II), (III) 
(we will mention them in detail in section 4) 
to be applied to the inverse of the 
non-resonant part of $^{16}$O propagator, $D_2(p)$, 
at $0\le E_\alpha \le 2.6$~MeV, where the experimental data are not available.
Then, we 
fit the parameters to the data 
and calculate the ANC of the $2_1^+$ state of $^{16}$O. 
We find that the parameters are fitted very well  
to the experimental phase shift data
for all the conditions (I), (II), (III), where 
the $\chi^2/N$ values for the parameter fit are less than one
or almost one for all cases;
the both small and large ANC values, 
depending on the choice of the conditions, 
are reproduced by using the fitted parameters.
Thus, it is not clear how one can pin down the value of ANC 
of the $2_1^+$ state of $^{16}$O 
from the phase shift data of elastic $\alpha$-$^{12}$C scattering.  
As already discussed in the literature, additional experimental input may
be necessary to determine the value of ANC of 
the $2_1^+$ state of $^{16}$O. 

The present work is organized as follows.
In section 2, an expression for the $S$ matrix is introduced
and an effective Lagrangian is presented,
and the elastic scattering amplitudes for $l=2$ 
are derived from the Lagrangian in section 3.
In section 4, we discuss that three conditions are imposed in low energy regions where
the experimental data do not exist 
and numerical results are obtained,
and results and discussion of this work are presented in section 5.

\vskip 2mm \noindent
{\bf 2. $S$ matrix and effective Lagrangian}

The $S$ matrix of elastic $\alpha$-$^{12}$C scattering for $d$-wave 
channel is given as
\bea
S_2 = e^{2i\delta_2}\,,
\eea
where $\delta_2$ is the phase shift for the $d$-wave elastic scattering
whose experimental values at 
$2.6 \le E_\alpha \le 6.62$~MeV are reported in Ref.~\cite{tetal-prc09}.
The scattering amplitude $\tilde{A}_2$
is related to the $S$ matrix as~\footnote{
There is a common factor difference between 
the expression of the amplitude $\tilde{A}_2$ and
the standard form of the amplitude $A_2$;  
$A_2 = \frac{10\pi}{\mu} e^{2i\sigma_2} \tilde{A}_2$
where $\sigma_2$ is the Coulomb phase shift for $l=2$,
$e^{2i\sigma_2} = \Gamma(3+i\eta)/\Gamma(3-i\eta)$ with
$\eta=\kappa/p$.
} 
\bea
S_2 = 1 + 2ip\tilde{A}_2\,.
\eea 
Because two resonant $2_2^+$ and $2_3^+$ states of $^{16}$O appear in the 
data at $E_\alpha(2_2^+) = 3.58$~MeV and $E_\alpha(2_3^+)=5.81$~MeV, 
respectively, we may decompose the phase shift $\delta_2$ as~\cite{g-prc09}
\bea
\delta_2 = \delta_2^{(nr)} + \delta_2^{(rs1)} + \delta_2^{(rs2)}\,,
\eea
where $\delta_2^{(nr)}$ is the phase shift for the background-like, 
non-resonant part 
and $\delta_2^{(rs1)}$ and $\delta_2^{(rs2)}$
are those for the resonant $2_2^+$
and $2_3^+$ states of $^{16}$O, respectively.
We assume that each of those phase shifts may have 
a relation to a corresponding scattering amplitude
as 
\bea
e^{2i\delta_2^{(ch)}} &=& 1 + 2ip\tilde{A}_2^{(ch)}\,,
\eea
where $ch(annel) = nr, rs1, rs2$, 
and $\tilde{A}_2^{(nr)}$, $\tilde{A}_2^{(rs1)}$, and $\tilde{A}_2^{(rs2)}$,
are the amplitudes for the non-resonant part, the first resonant part,
and the second resonant part of the amplitudes, which will be constructed 
from the effective Lagrangian in below. 
Thus, the total amplitude $\tilde{A}_2$ for the nuclear reaction part  
in terms of the three amplitudes, $\tilde{A}_2^{(nr)}$, $\tilde{A}_2^{(rs1)}$,
$\tilde{A}_2^{(rs2)}$, is 
\bea
\tilde{A}_2 &=&
\tilde{A}_2^{(nr)} 
+ e^{2i\delta_2^{(nr)}} \tilde{A}_2^{(rs1)} 
+ e^{2i(\delta_2^{(nr)}+\delta_2^{(rs1)})} \tilde{A}_2^{(rs2)} \,.
\label{eq;tldA2}
\eea

An effective Lagrangian to derive the scattering amplitude
for the $d$-wave elastic $\alpha$-$^{12}$C scattering 
at low energies including the sub-threshold $2_1^+$ state of $^{16}$O and
the resonant $2_2^+$ and $2_3^+$ states of $^{16}$O 
may be written 
as~\cite{a-prc18,a-epja16,a-jkps18,a-epja07}
\bea
{\cal L} &=& 
\phi_\alpha^\dagger \left(
iD_0 
+\frac{\vec{D}^2}{2m_\alpha}
\right) \phi_\alpha
+ \phi_C^\dagger\left(
iD_0
+ \frac{\vec{D}^2}{2m_C}
\right)\phi_C
\nnb \\ && +
\sum_{k=0}^3 
C_{k}^{(nr)}d_{(nr)ij}^\dagger 
\left[
iD_0 
+ \frac{\vec{D}^2}{2(m_\alpha+m_C)}
\right]^k d_{(nr)ij}
\nnb \\ && 
- y_{(nr)}\left[
d_{(nr)ij}^\dagger(\phi_\alpha O_{2,ij}  \phi_C)
+ (\phi_\alpha O_{2,ij}  \phi_C)^\dagger d_{(nr)ij}
\right] 
\nnb \\ && +
\sum_{N=1}^2 \sum_{k=0}^3 
C_{k}^{(rsN)}d_{(rsN)ij}^\dagger 
\left[
iD_0 
+ \frac{\vec{D}^2}{2(m_\alpha+m_C)}
\right]^k d_{(rsN)ij}
\nnb \\ &&
- \sum_{N=1}^2y_{(rsN)}\left[
d_{(rsN)ij}^\dagger(\phi_\alpha O_{2,ij}  \phi_C)
+ (\phi_\alpha O_{2,ij}  \phi_C)^\dagger d_{(rsN)ij}
\right]\,, 
\eea
where $\phi_\alpha$ ($m_\alpha$) and 
$\phi_C$ ($m_C$) are scalar fields (masses) of $\alpha$ and $^{12}$C, 
respectively.  
$D^\mu$ is a covariant derivative,
$D^\mu = \partial^\mu + i {\cal Q}A^\mu$ where
${\cal Q}$ is the charge operator and $A^\mu$ is the photon field.
$d_{(nr)ij}$ and $d_{(rsN)ij}$ with $N=1,2$
are the composite fields of $^{16}$O 
consisting of 
$\alpha$ and $^{12}$C fields for $l=2$ 
for the non-resonant part ($nr$) representing 
the sub-threshold $2_1^+$ state of $^{16}$O,
and the two resonant parts ($rs1$) and ($rs2$) 
representing 
the first and second resonant $2_2^+$ and
$2_3^+$ states 
of $^{16}$O, 
respectively, which are introduced 
for perturbative expansion around the unitary 
limit~\cite{b-pr49,k-npb97,bs-npa01,ah-prc05}.  
The coupling constants of the non-resonant part of the amplitude, 
$C_k^{(nr)}$ with $k=0,1,2,3$, correspond to 
the effective range parameters of elastic $\alpha$-$^{12}$C
scattering while the first coupling constant $C_0^{(nr)}$ is fixed
by using the binding energy of the sub-threshold $2_1^+$ state of $^{16}$O
and the other parameters are fitted to the experimental phase shift data
with other parameters appearing in the $S$ matrix.
The coupling constants of the resonant parts, 
$C_k^{(rs1)}$, and $C_k^{(rs2)}$ 
with $k=0,1,2,3$,
are rewritten in terms of the first two terms, 
$C_{0}^{(rs1)}$ and $C_{1}^{(rs1)}$ 
as well as $C_{0}^{(rs2)}$ and $C_{1}^{(rs2)}$, 
by using the resonant energies and widths for the 
resonant $2_2^+$ and $2_3^+$ 
states of $^{16}$O, respectively. 
The third and fourth parameters for the resonant $2_3^+$ state, 
$C_{2}^{(rs2)}$ and
$C_{3}^{(rs2)}$  
are fitted to the phase shift data while  
we set 
$C_{2}^{(rs1)}=
C_{3}^{(rs1)}=0$ 
for the first resonant $2_2^+$ state of $^{16}$O.   
The coupling constants $y_{(nr)}$ and $y_{(rsN)}$ with $N=1,2$ are
uniquely defined but convention-dependent~\cite{g-npa04}.
We take the convenient choice; 
$y_{(nr)} = y_{(rs1)} =y_{(rs2)} = \sqrt{2\pi/\mu}$
where $\mu$ is the reduced mass of $\alpha$ and 
$^{12}$C, 
as a trade-off for a complicated redefinition of the respective 
composite fields. 
\begin{figure}[t]
\begin{center}
\resizebox{0.7\textwidth}{!}{
  \includegraphics{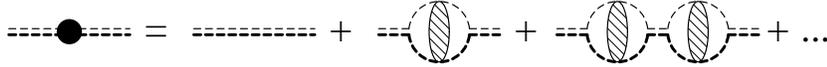}
}
\caption{
Diagrams for dressed $^{16}$O propagator.
A thick (thin) dashed line represents a propagator of $^{12}$C ($\alpha$),
and a thick and thin double dashed line with and without a filled circle
represent a dressed and bare $^{16}$O propagator, respectively.
A shaded blob represents
a set of diagrams consisting of all possible one-potential-photon-exchange
diagrams up to infinite order and no potential-photon-exchange one.
}
\label{fig;propagator}       
\end{center}
\end{figure}
\begin{figure}[t]
\begin{center}
\resizebox{0.2\textwidth}{!}{
 \includegraphics{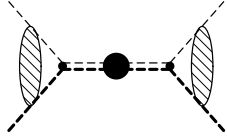}
}
\caption{
Diagram of the scattering amplitude.
See the caption of Fig.~\ref{fig;propagator} as well.
}
\label{fig;scattering_amplitude}      
\end{center}
\end{figure}

The scattering amplitudes, $\tilde{A}_2^{(nr)}$ and 
$\tilde{A}_2^{(rsN)}$ with $N=1,2$, 
are calculated from the diagrams depicted in 
Figs.~\ref{fig;propagator} and \ref{fig;scattering_amplitude}.
Here, the bubble diagrams are summed up to the infinite order 
in Fig.~\ref{fig;propagator}.
For the non-resonant part of the amplitude, 
we treat it non-perturbatively because of the study for 
the ANC of $2_1^+$ state of $^{16}$O at its binding energy.
For the resonant parts of the amplitude
(for our case, they are classified as narrow resonances 
because of $\Gamma_r \ll E_r$~\cite{hfvk-epja21}), 
the counting rules of resonant states are carefully studied by 
Gelman~\cite{g-prc09} and Habashi et al.~\cite{hfvk-epja21}.
The energy range of phase shift
data covers the two resonant states, and at the vicinities of the resonant
energies we should have the amplitudes for which the bubble diagrams 
are summed up to the infinite order. 
While at the off-resonant energy regions, 
one can expand the resonant amplitudes perturbatively and the
$d_{(nr)ij}$ and $d_{(rsN)ij}$ with $N=1,2$ fields may start mixing 
through the bubble diagram for corrections at higher orders. 
We keep the summed amplitudes for the resonant states 
as leading contributions
and ignore the field mixing in the present study.  

\vskip 2mm \noindent
{\bf 3. Scattering amplitudes}

For the non-resonant amplitude $\tilde{A}_2^{(nr)}$, 
we have~\cite{a-prc18,a-epja16} 
\bea
\tilde{A}^{(nr)}_2 &=&
\frac{C_\eta^2 W_2(p)}{
K_2(p) 
-2\kappa H_2(p)
}
\label{eq;Aer2}
\label{eq;A2_nr}
\,,
\eea
where the function $C_\eta^2 W_2(p)$ in the numerator of the amplitude
is calculated from the initial and final state Coulomb interactions 
in Fig.~\ref{fig;scattering_amplitude}; $p$ is the magnitude of 
relative momentum of the $\alpha$-$^{12}$C system in the center of mass
frame, $p=\sqrt{2\mu E}$. Thus, one has
\bea
C_\eta^2 &=& \frac{2\pi\eta}{\exp(2\pi\eta)-1}\,,
\\
W_2(p) &=& \frac14(\kappa^2+p^2)(\kappa^2+4p^2)\,,
\eea
where $\eta=\kappa/p$: 
$\kappa$ is the inverse of the Bohr radius,
$\kappa = Z_2Z_6 \alpha_E\mu$, where $Z_n$ are the number of protons
of the nuclei, $Z_2=2$ and $Z_6=6$, 
and $\alpha_E$ is the fine structure constant.
The function $-2\kappa H_2(p)$ in the denominator of the amplitude
is the Coulomb self-energy term which is calculated 
from the loop diagram in Fig.~\ref{fig;propagator}, and one has 
\bea
H_2(p) &=& W_2(p) H(\eta)\, ,
\ \ \ 
H(\eta) = \psi(i\eta) + \frac{1}{2i\eta} -\log(i\eta)\,,
\eea
where $\psi(z)$ is the digamma function. 
The nuclear interaction is represented 
in terms of the effective range parameters 
in the function $K_2(p)$ in
the denominator of the amplitude in Eq.~(\ref{eq;Aer2}).
As discussed in Ref.~\cite{a-prc18}, large and significant contributions
to the series of effective range expansion,
compared to that evaluated from the phase shift data at the lowest energy
of the data, $E_\alpha=2.6$~MeV, 
appear from the Coulomb self-energy term, $-2\kappa H_2(p)$. To 
subtract those contributions, we include the effective range terms up to
$p^6$ order as counterterms. For more detail, see the appendix.
Thus, we have  
\bea
K_2(p) &=&
-\frac{1}{a_2}
+\frac12 r_2p^2
-\frac14 P_2 p^4
+Q_2 p^6\,,
\eea
where $a_2$, $r_2$, $P_2$, $Q_2$ are effective range parameters. 

Now we fix a parameter among the four effective range parameters,
$a_2$, $r_2$, $P_2$, and $Q_2$, by using the condition that 
the inverse of the scattering amplitude $\tilde{A}_2^{(nr)}$ vanishes at 
the binding energy of the sub-threshold $2_1^+$ state of $^{16}$O. 
Thus, the denominator of the scattering amplitude,
\bea 
D_2(p) = K_2(p) -2\kappa H_2(p) \,,  
\label{eq;binding_pole}
\eea
vanishes at $p=i\gamma$ where $\gamma$ is the binding momentum 
of the $2_1^+$ state of $^{16}$O; 
$\gamma = \sqrt{2\mu B_{2}}$ where $B_2$ 
is the binding energy of the $2_1^+$ state of $^{16}$O
from the $\alpha$-$^{12}$C breakup threshold.
Using the condition, $D_2(i\gamma)=0$, 
we fix the effective range parameter
$a_2$ as
\bea
-\frac{1}{a_2} &=& 
\frac12\gamma^2 r_2
+\frac14\gamma^4 P_2
+\gamma^6Q_2 
+ 2\kappa H_2(i\gamma)\,. 
\label{eq;a2}
\eea
Using the relation in Eq.~(\ref{eq;a2}),
we rewrite the denominator of the amplitude $D_2(p)$ as
\bea
D_2(p) &=& 
\frac12 r_2 (\gamma^2+p^2)
+\frac14 P_2 (\gamma^4-p^4)
+ Q_2 (\gamma^6 + p^6)
+ 2\kappa \left[ H_2(i\gamma) - H_2(p)
\right]\,,
\eea
where we have three constants, $r_2$, $P_2$, $Q_2$, 
in the function $D_2(p)$ 
for the non-resonant amplitude $\tilde{A}_2^{(nr)}$,
which are fitted to the phase shift data. 

For the elastic scattering amplitudes for the resonant $2_2^+$ and
$2_3^+$ states of $^{16}$O, 
we may first have those amplitudes as the same expression of the non-resonant
amplitude in Eq.~(\ref{eq;Aer2})
in terms of the effective range expansion as
\bea
\tilde{A}^{(rsN)}_2 &=&
\frac{C_\eta^2W_2(p)}{
K^{(rsN)}_2(p)
-2\kappa H_2(p)
}
\label{eq;A2_rs}
\,,
\eea
with $N=1,2$, which correspond to the first and second 
resonant $2_2^+$ and $2_3^+$ states of $^{16}$O, respectively, and 
\bea
K^{(rsN)}_2(p) &=&
-\frac{1}{a_2^{(rsN)}}
+\frac12 r_2^{(rsN)} p^2
-\frac14 P_2^{(rsN)}  p^4
+Q_2^{(rsN)} p^6\,.
\label{eq;K2rsN}
\eea

We now introduce 
the expansion around the resonant energies in the denominator
of the scattering amplitudes~\cite{hhvk-npa08}.
Thus, we rewrite the amplitudes as  
\bea
\tilde{A}_2^{(rsN)} &=&
- 
\frac{1}{p}
\frac{\frac12\Gamma^{(rsN)}(E) }{E-E_r^{(rsN)} 
+ R^{(rsN)}(E) + i\frac12\Gamma^{(rsN)}(E)}\,,
\label{eq;A2_rsN}
\eea
with
\bea
\Gamma^{(rsN)}(E) &=& \Gamma_r^{(rsN)}
\frac{pW_2(p)C_\eta^2}{
p_rW_2(p_r)C_{\eta_r}^2}\,,
\\
R^{(rs2)}(E) &=& a(E-E_r^{(rs2)})^2 + b(E-E_r^{(rs2)})^3 
\,,
\label{eq;R}
\eea
and 
$R^{(rs1)}(E) = 0$, where 
\bea
a &=& \frac12 Z_r 
\left(
2P_2^{(rs2)} \mu^2
-48Q_2^{(rs2)} \mu^3 E_r^{(rs2)}
+ 2\kappa Re 
\left. 
\frac{\partial^2 H_2}{\partial E^2}
\right|_{E=E_r^{(rs2)}}
\right) \,,
\\
b &=& \frac16 Z_r
\left(
-48Q_2^{(rs2)} \mu^3 
+ 2\kappa \left. Re
\frac{\partial^3 H_2}{\partial E^3}
\right|_{E=E_r^{(rs2)}}
\right)\,,
\\
Z_r^{-1} &=&
Re \left.
\frac{\partial}{\partial E} D_2^{(rs2)}(E)\right|_{E=E_r^{(rs2)}}\,,
\ \ \ \
Z_r = 
\frac{\Gamma_r^{(rs2)}}{
2p_rW_2(p_r)C_{\eta_r}^2
}\,.
\eea
In the above equations,
$E_r^{(rsN)}$ and $\Gamma_r^{(rsN)}$ with $N=1,2$ are 
the energies and the widths of the
resonant states of $^{16}$O, and $p_r$ are the resonant momenta,
$p_r=\sqrt{2\mu E_r^{(rsN)}}$, 
which also appear in $\eta_r$ as $\eta_r=\kappa/p_r$.

Using the expression of the amplitudes 
in Eqs.~(\ref{eq;A2_nr}) and (\ref{eq;A2_rsN}), 
we have the $S$ matrix as
\bea
e^{2i\delta_2} &=& 
\frac{K_2(p) - 2\kappa Re H_2(p) + ipC_\eta^2W_2(p)}
     {K_2(p) - 2\kappa Re H_2(p) - ipC_\eta^2W_2(p)}
\nnb \\ && \times
\frac{E - E_r^{(rs1)} - i\frac12\Gamma^{(rs1)}(E)}
     {E - E_r^{(rs1)} + i\frac12\Gamma^{(rs1)}(E)}
\
\frac{E - E_r^{(rs2)} + R^{(rs2)}(E) - i\frac12\Gamma^{(rs2)}(E)}
     {E - E_r^{(rs2)} + R^{(rs2)}(E) + i\frac12\Gamma^{(rs2)}(E)}\,.
\label{eq;exp2idel2} 
\eea
We will fit the parameters in the $S$ matrix to the phase shift data
by using the sine function of the phase shift,
$f=\sin(\delta_2)$,
in the next section; the scattering cross section is proportional to 
$\sin^2(\delta_2)$. 
For the study of the ANC, at the small energy region, 
the exponential factors in Eq.~(\ref{eq;tldA2}) 
almost become one
due to the Gamow factor in $C_\eta^2$, and the amplitude becomes
\bea
\tilde{A}_2 &=&
\tilde{A}_2^{(nr)} 
+ \tilde{A}_2^{(rs1)} 
+ \tilde{A}_2^{(rs2)} 
+ O(C_\eta^4) \,,  
\eea
where the pole at the $2_1^+$ state of $^{16}$O exists
in $\tilde{A}_2^{(nr)}$,
the ANC $|C_b|$ for the $2_1^+$ state of $^{16}$O is calculated 
by using a formula~\cite{ir-prc84}
\bea
|C_b| &=& \frac12 \gamma^2 \Gamma(3+\kappa/\gamma)
\left[
\left.
- \frac{\partial D_2(p)}{\partial p^2}
\right|_{p^2 = -\gamma^2}
\right]^{-1/2}\,,
\label{eq;Cb}
\eea
where $\Gamma(z)$ is the gamma function.

\vskip 2mm \noindent
{\bf 4. Numerical results}

Nine parameters, $\theta = \{ 
r_2, P_2, Q_2, 
E_r^{(rs1)}, \Gamma_r^{(rs1)}, 
E_r^{(rs2)}, \Gamma_r^{(rs2)}, 
a,b \}$, appear in the $S$ matrix, where
$r_2$, $P_2$, and $Q_2$ are the effective range parameters, which reproduce 
the binding energy of the sub-threshold $2_1^+$ state of $^{16}$O 
in $\tilde{A}_2^{(nr)}$, $E_r^{(rs1)}$ and $\Gamma_r^{(rs1)}$ are the 
resonant energy and width of the $2_2^+$ state of $^{16}$O 
in $\tilde{A}_2^{(rs1)}$, and $E_r^{(rs2)}$ and $\Gamma_r^{(rs2)}$ are
those of the $2_3^+$ state of $^{16}$O in $\tilde{A}_2^{(rs2)}$.
While $a$ and $b$ are the coefficients of higher order terms 
$(E-E_r^{(rs2)})^n$ with $n=2,3$ in the $R^{(rs2)}(E)$ function, respectively,
obtained expanding the denominator of $\tilde{A}_2^{(rs2)}$ around the resonant
energy, $E=E_r^{(rs2)}$.
We treat the parameters $a$ and $b$ as independent parameters
for the sake of simplicity though they are functions of 
$P_2^{(rs2)}$, $Q_2^{(rs2)}$ and $E_r^{(rs2)}$.
As mentioned in the introduction, we also study the inclusion of 
$2_4^+$ state of $^{16}$O; this can be done straightforwardly. 
The expression for the $2_4^+$ state of $^{16}$O 
in the $S$ matrix is the same 
as that for the $2_2^+$ state of $^{16}$O in Eq.~(\ref{eq;exp2idel2}), 
but we use the fixed experimental values of the energy and width, 
$E_r^{(rs3)}$ and $\Gamma_r^{(rs3)}$.
Thus, we employ two expressions of the $S$ matrix for the parameter fit; 
one is the $S$ matrix given in Eq.~(\ref{eq;exp2idel2}),
and the other is that including the contribution from 
the $2_4^+$ state of $^{16}$O.   
We fit the nine parameters in the $S$ matrices 
(by using the fitting function $f=\sin(\delta_2)$, as mentioned above) 
to the phase shift data $\delta_2$ of the 
elastic $\alpha$-$^{12}$C scattering for $d$-wave channel 
reported by Tischhauser et al.~\cite{tetal-prc09}, 
by employing an Markov chain Monte Carlo (MCMC) program~\cite{emcee}.  

In addition, to investigate the difference between the large ANC values
obtained from the $\alpha$ transfer reactions and the small ANC values
from the effective range expansion,
we impose three different conditions to the inverse of 
the $^{16}$O propagator, $D_2(p)$, of the non-resonant amplitude
at the very low $\alpha$ energy region, $0< E_\alpha <2.6$~MeV,
where the experimental data are not available. 
We note that the function $D_2(p)$ should be negative at $0<E_\alpha <2.6$~MeV.
Those three conditions are 
\begin{itemize}
\item (I) $D_2(p)<0$, 

\item (II) $D_2(p_{i+1}) < D_2(p_i)$, 

\item (III) $\left. \frac{dD_2}{dp^2}\right|_{p=p_{i+1}} < 
\left.\frac{dD_2(p)}{dp^2}\right|_{p=p_i}$, 
\end{itemize}
where $p_{i+1}>p_i$ and $p=\sqrt{2\mu E} = \sqrt{1.5\mu E_\alpha}$. 
The first condition (I) is always required because there should be no zeros
for no resonant states of $^{16}$O at the very low energy region. 
The second condition (II) is modest and 
the condition (III) can reproduce 
the large ANC values reported from the $\alpha$ transfer reactions
as we will see below.  
We note that those conditions are easily included 
as conditions in the prior distribution
with the MCMC method. 

For the parameter fit, we treat the set of the parameters, 
$\theta = \{
r_2,P_2,Q_2, 
E_r^{(rs1)}, \Gamma_r^{(rs1)},
E_r^{(rs2)}, \Gamma_r^{(rs2)},
a,b\}$ 
as free parameters 
while the values of $r_2$, $P_2$, $Q_2$ are constrained due to the conditions
(I), (II), (III) in the inverse of the $^{16}$O propagator, $D_2(p)$.
For the initial values of the parameters, $r_2$, $P_2$, $Q_2$,
we employ the values reported by 
Sparenberg, Capel, and Baye~\cite{scb-jp11};
$r_2=0.1580$~fm$^{-3}$,
$P_2= -1.041$~fm$^{-1}$,
$Q_2 = 0.1411$~fm, which lead to a large ANC value,
$|C_b|=13.8\times 10^4$~fm$^{-1/2}$. 
We also choose the initial values of $a$ and $b$ as 
$a=0.1 (0.5)$~MeV$^{-1}$ and $b=1.1 (0.5)$~MeV$^{-2}$
for the $S$ matrix without (with) the $2_4^+$ state of $^{16}$O. 
While we use the center values of the experimental resonant energies 
and widths of the 
$2_2^+$ and $2_3^+$ states of $^{16}$O; 
$E_r^{(rs1)} = 2.6826(5)$~MeV, $\Gamma_r^{(rs1)} = 0.625(100)$~keV,
$E_r^{(rs2)} = 4.358(4)$~MeV, $\Gamma_r^{(rs2)} = 71(3)$~keV~\cite{twc-npa93}, 
as initial values for the parameter fit.
We use the fixed experimental values for the energy and width of the 
$2_4^+$ state of $^{16}$O as $E_r^{(rs3)}=5.858$~MeV and 
$\Gamma_r^{(rs3)}=150$~keV~\cite{twc-npa93}. 
Details for the parameter fit and the calculation of the error bars 
can be found in our previous work~\cite{ya-jkps19}.

\begin{table}[t]
\begin{center}
\begin{tabular}{c || c c c } 
       & (I) & (II) & (III) \cr \hline \hline
$r_2$ (fm$^{-3}$)  &  0.137(4) & 0.150(4) & 0.159(4) \cr 
$P_2$ (fm$^{-1}$) & $-1.36(5)$ & $-1.18(4)$  & $-1.07(4)$ \cr
$Q_2$ (fm) & $0.013(16)$ & $0.075(12)$ & $0.112(11)$ \cr \hline
$E_r^{(rs1)}$ (MeV) & 2.68308(5) & 2.68309(5) & 2.68309(5) \cr
$\Gamma_r^{(rs1)}$ (keV) & 0.75(2) & 0.74(2) & 0.74(2) \cr \hline
$E_r^{(rs2)}$ (MeV) & 4.3549(1) & 4.3545(1) & 4.3544(1) \cr
$\Gamma_r^{(rs2)}$ (keV) & 74.66(3) & 74.60(3) & 74.57(3) \cr
$a$ (MeV$^{-1}$) & $0.20(6)$ & $0.47(6)$ & $0.59(5)$ \cr
$b$ (MeV$^{-2}$) & $0.94(5)$ & $1.13(8)$ & $1.39(10)$ \cr 
\hline \hline
$|C_b|$ (fm$^{-1/2}$) & $2.0(2)\times 10^4$ & 
 $3.2(6)\times 10^4$ & $11(26)\times 10^4$  \cr
$\chi^2/N$ &  0.74 & 0.86 & 1.13 \cr 
\hline
\end{tabular}
\caption{
Values and errors of nine parameters
$\{
r_2, P_2, Q_2, 
E_r^{(rs1)}, \Gamma_r^{(rs1)}, 
E_r^{(rs2)}, \Gamma_r^{(rs2)}, 
a, b\}$ 
in the $S$ matrix in Eq.~(\ref{eq;exp2idel2})
fitted to the experimental phase shift $\delta_2$
of elastic $\alpha$-$^{12}$C scattering for $l=2$ 
using the conditions (I), (II), (III) at the low energy region.
The ANC, $|C_b|$,  for the $2_1^+$ 
states of $^{16}$O are calculated by using the fitted values 
of the parameters, and values of $\chi^2/N$, where $N$ is the number
of data, for the fit are presented in the last row of the table. 
}
\label{table;parameters_A}
\end{center}
\end{table}

\begin{table}[t]
\begin{center}
\begin{tabular}{c || c c c } 
       & (I) & (II) & (III) \cr \hline \hline
$r_2$ (fm$^{-3}$)  &  0.149(4) & 0.152(4) & 0.159(3) \cr 
$P_2$ (fm$^{-1}$) & $-1.19(5)$ & $-1.16(4)$  & $-1.07(3)$ \cr
$Q_2$ (fm) & $0.081(16)$ & $0.090(14)$ & $0.121(9)$ \cr \hline
$E_r^{(rs1)}$ (MeV) & 2.68308(5) & 2.68308(5) & 2.68309(5) \cr
$\Gamma_r^{(rs1)}$ (keV) & 0.75(2) & 0.75(2) & 0.75(2) \cr \hline
$E_r^{(rs2)}$ (MeV) & 4.3545(2) & 4.3545(1) & 4.3542(1) \cr
$\Gamma_r^{(rs2)}$ (keV) & 74.61(3) & 74.59(3) & 74.55(3) \cr
$a$ (MeV$^{-1}$) & $0.46(12)$ & $0.51(10)$ & $0.71(6)$ \cr
$b$ (MeV$^{-2}$) & $0.47(9)$ & $0.53(10)$ & $0.81(11)$ \cr 
\hline \hline
$|C_b|$ (fm$^{-1/2}$) & $3.1(6)\times 10^4$ & 
 $3.6(9)\times 10^4$ & $13(30)\times 10^4$  \cr
$\chi^2/N$ &  0.66 & 0.69 & 0.73 \cr 
\hline
\end{tabular}
\caption{
Values and errors of nine parameters
$\{
r_2, P_2, Q_2, 
E_r^{(rs1)}, \Gamma_r^{(rs1)}, 
E_r^{(rs2)}, \Gamma_r^{(rs2)}, 
a, b\}$ 
in the $S$ matrix, which contains 
the contribution from the $2_4^+$ state of $^{16}$O,
fitted to the experimental phase shift $\delta_2$
of elastic $\alpha$-$^{12}$C scattering for $l=2$ 
using the conditions (I), (II), (III) at the low energy region.
See the caption of Table \ref{table;parameters_A} as well. 
}
\label{table;parameters_B}
\end{center}
\end{table}

We are now in a position to discuss the numerical results of the present work. 
In Tables 
\ref{table;parameters_A} and 
\ref{table;parameters_B}, 
we present the values and errors of the parameters 
in the $S$ matrices without and with the contribution from the $2_4^+$ state
of $^{16}$O, respectively,
fitted to the phase shift data 
imposing the three conditions (I), (II), (III). 
We also include in the tables the values and errors of the ANC, 
$|C_b|$, of the $2_1^+$ state of $^{16}$O and the $\chi^2/N$ values 
where $N$ is the number of data, $N=354$.

In the tables, one can see that
the values of $\chi^2/N$ increase as the conditions (I), (II), (III) 
are orderly changed.
This is caused by tighter restrictions being applied to 
the effective range parameters, $r_2$, $P_2$, $Q_2$,
as altering the conditions (I), (II), (III) in order.  
Nevertheless, all values of the parameters may be regarded to be 
fitted very well to the phase shift data 
because the values of $\chi^2/N$ 
are smaller than one or almost one
whereas the inclusion of the $2_4^+$ state of $^{16}$O makes the 
$\chi^2/N$ values in Table~\ref{table;parameters_B} smaller than those in 
Table~\ref{table;parameters_A}.  

The fitted parameters in the tables can be parted into two groups 
depending on sensitivities to the conditions (I), (II), (III).
One can easily see that the energies and widths of the resonant states
of $^{16}$O are insensitive to the conditions; they are determined by the 
significant resonant peaks. We find that the fitted values of $E_r^{(rs1)}$,
$\Gamma_r^{(rs1)}$, $E_r^{(rs2)}$, $\Gamma_r^{(rs2)}$, basically 
agree with the experimental values~\cite{twc-npa93}.  
The fitted values of $r_2$, $P_2$, $Q_2$, $a$, $b$ are sensitive to the 
conditions (I), (II), (III); they are fitted to the non-resonant
part of the phase shift data at the energy regions 
below the first resonant state and between the two resonant states. 
One may notice that the values of $b$ in Table~\ref{table;parameters_B} 
are remarkably smaller than those in Table~\ref{table;parameters_A},
apparently due to the effect of the $2_4^+$ state of $^{16}$O.
We can reproduce the small and large values of the ANC, $|C_b|$,
of the $2_1^+$ state of $^{16}$O by using the fitted values of 
the effective range parameters, $r_2$, $P_2$, $Q_2$ for the conditions
(I), (II), (III), where the effective 
range parameters for the condition (III) reasonably agree well with those 
reported by Sparenberg, Capel, and Baye for the large ANC value. 
In addition, a very large error bar appears in the large ANC value, so 
the small and large values of the ANC in fact agree with each other 
within the error bars.
In the present study, therefore, 
it is not easy to pin down which of the ANC values
is correct because of the error bars of the ANC values 
and the $\chi^2/N$ values discussed above.

\begin{figure}[t]
\begin{center}
  \includegraphics[width=12cm]{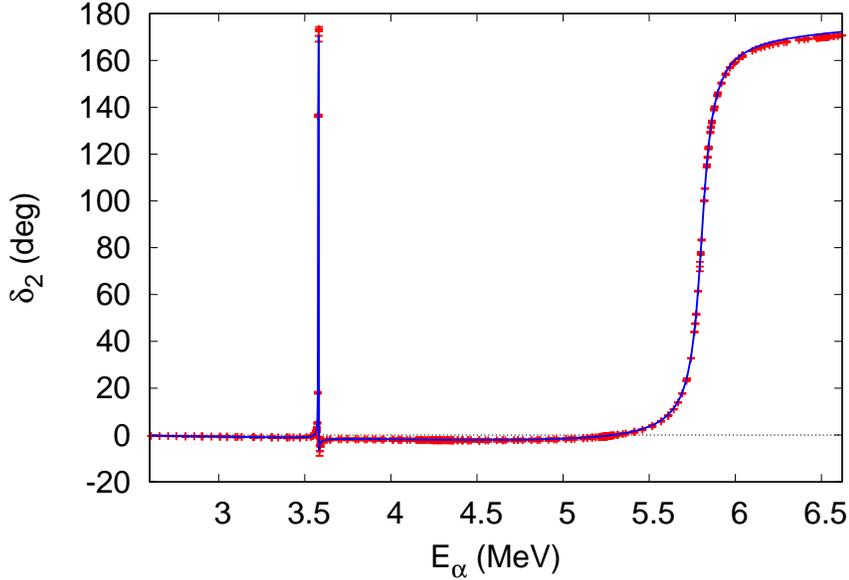}
\caption{
Phase shift $\delta_2$ of elastic $\alpha$-$^{12}$C scattering
for $d$-wave channel 
as a function of 
$E_\alpha$ calculated by using the fitted values of the parameters 
in the (I) column in Table~\ref{table;parameters_A}.
Experimental data are included in the figure as well. 
}
\label{fig;del2}       
\end{center}
\end{figure}

\begin{figure}[t]
\begin{center}
  \includegraphics[width=12cm]{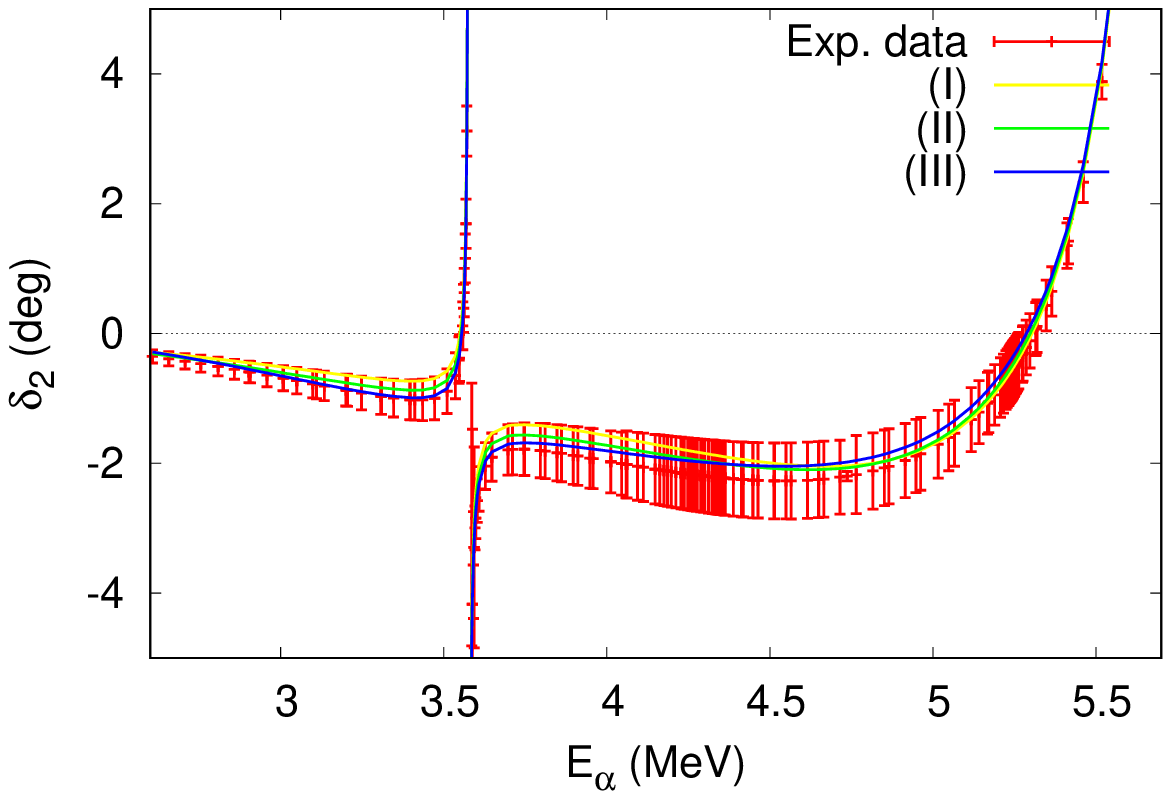}
\caption{
Phase shift $\delta_2$ of elastic $\alpha$-$^{12}$C scattering 
for $d$-wave channel at the small phase shift region 
as a function of 
$E_\alpha$ calculated by using 
the fitted values of the parameters in 
Table~\ref{table;parameters_A}.
Experimental data are included in the figure as well. 
}
\label{fig;del2_small_A}       
\end{center}
\end{figure}

\begin{figure}[t]
\begin{center}
  \includegraphics[width=12cm]{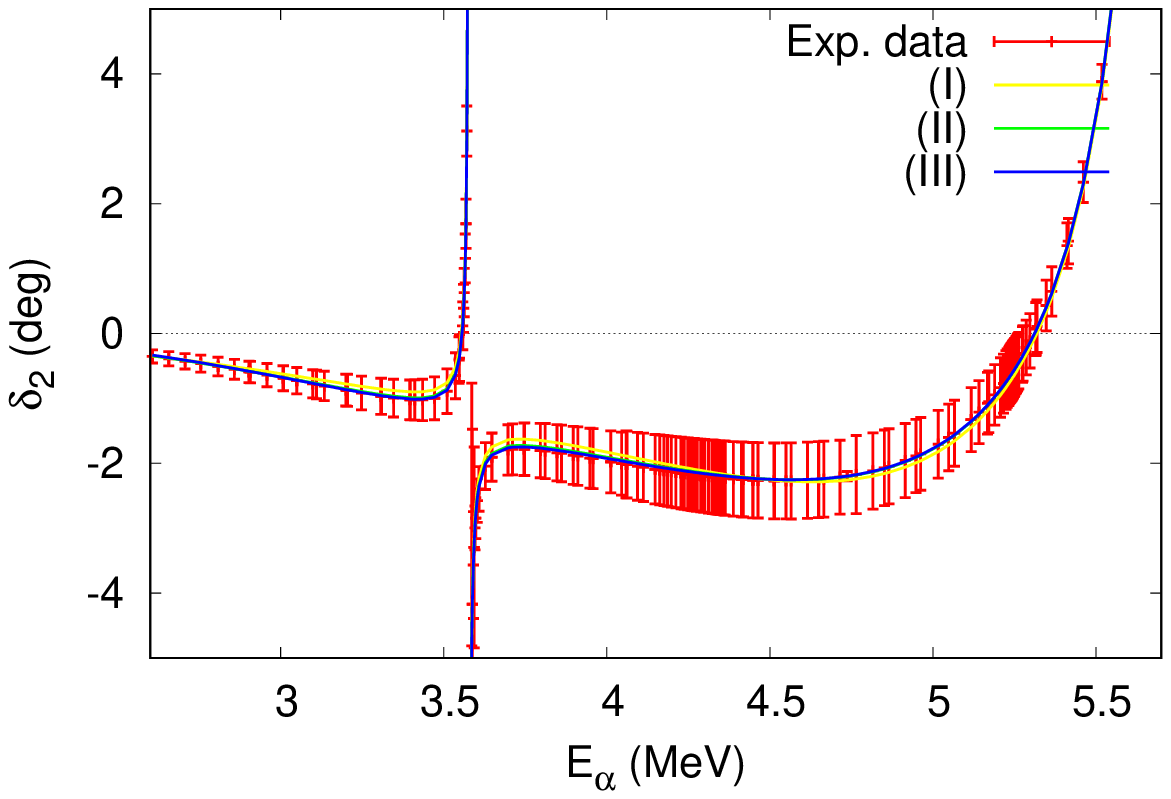}
\caption{
Phase shift $\delta_2$ of elastic $\alpha$-$^{12}$C scattering 
for $d$-wave channel at the small phase shift region
as a function of 
$E_\alpha$ calculated by using 
the fitted values of the parameters in 
Table~\ref{table;parameters_B}.
Experimental data 
are included in the figure as well. 
}
\label{fig;del2_small_B}       
\end{center}
\end{figure}
In Fig.~\ref{fig;del2}, we plot the phase shift $\delta_2$ of 
elastic $\alpha$-$^{12}$C scattering for $d$-wave channel
calculated by using the fitted values of the parameters in the (I) column
in Table~\ref{table;parameters_A}
as a function of the $\alpha$ energy, $E_\alpha$, and 
the phase shift data are also included in the figure.  
This is a typical figure for almost all of the cases; two peaks
of the resonant $2_2^+$ and $2_3^+$ states of $^{16}$O are well reproduced.
The differences due to the values of the fitted parameters are hardly seen
at the low energy tail and the energies between the two resonant states.

In Figs.~\ref{fig;del2_small_A} and \ref{fig;del2_small_B}, 
we plot the curves of the phase shift $\delta_2$ 
for the small $\delta_2$ values, $-5^\circ \le \delta_2 \le 5^\circ$,
which are
calculated by using the values of the parameters in the $S$ matrices 
without and with the $2_4^+$ state of $^{16}$O, respectively,
fitted to the phase shift data 
for the low energy tail and the energies between the two resonant states. 
The phase shift data are included in the figures as well. 
One can see that the calculated curves 
agree very well within the error bars of the data
as discussed above that the $\chi^2/N$ values are less than one or almost
one for all the cases.  

\begin{figure}[t]
\begin{center}
  \includegraphics[width=12cm]{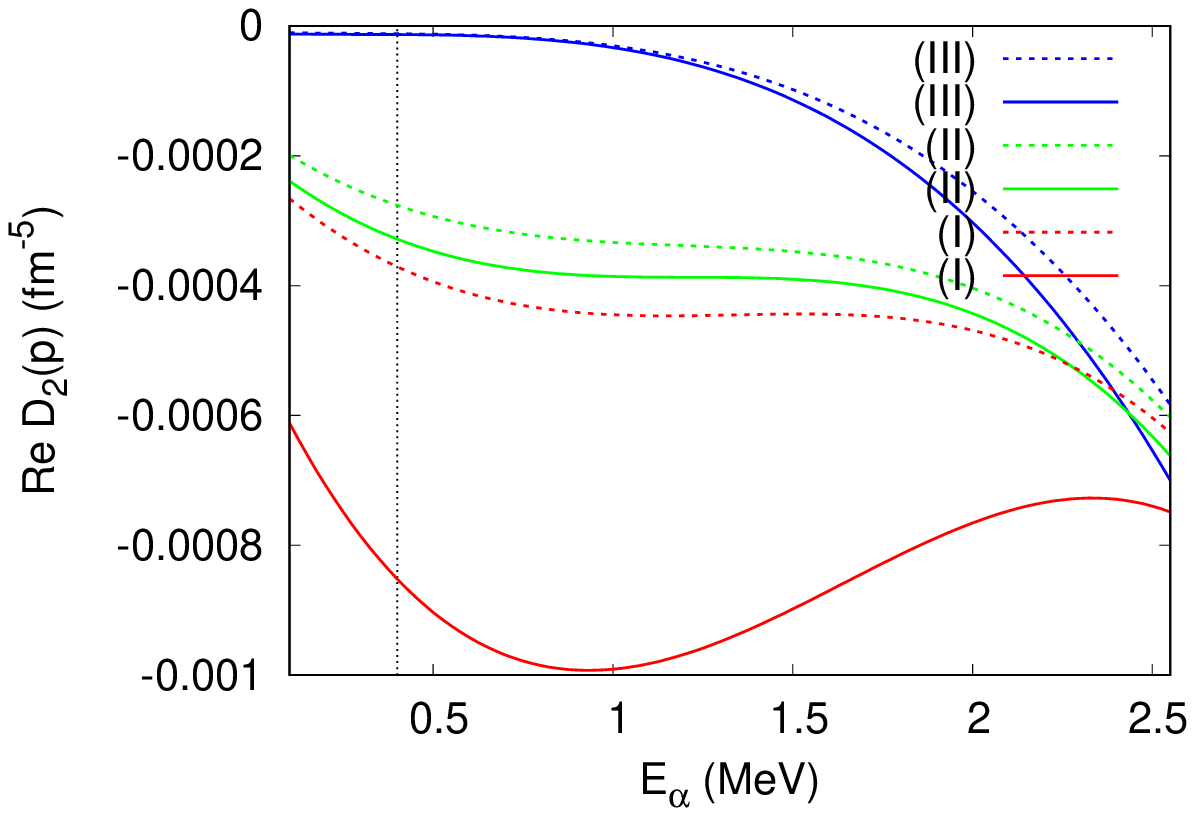}
\caption{
Real part of inverse of dressed $^{16}$O propagator for $l=2$, $D_2(p)$, 
calculated by using the values of the parameters with the conditions
(I), (II), (III) 
in Table~\ref{table;parameters_A} (curves) 
and in Table~\ref{table;parameters_B} (dashed curves)  
as a function of $E_\alpha$ at the energies where the experimental 
data are not available.  A dotted vertical line at $E_\alpha = 0.4$~MeV is
also drawn in the figure. 
}
\label{fig;ReD2}       
\end{center}
\end{figure}
In Fig.~\ref{fig;ReD2},
we plot the real part of $D_2(p)$ calculated by using 
the values of the parameters
in Table~\ref{table;parameters_A} (as curves)
and in Table~\ref{table;parameters_B} (as dashed curves) 
for the conditions (I), (II), (III)
as a function of $E_\alpha$ at 
$0\le E_\alpha\le 2.6$~MeV, where the phase shift data are not available.
One can see that those curves satisfy the conditions (I), (II), (III) and 
go through the quite different values of $ReD_2(p)$ 
at an energy range, $0\le E_\alpha \le 1.5$~MeV
whereas the inclusion of the $2_4^+$ state of $^{16}$O makes the curve of (I)
less deviated from the other curves. 
When one calculates the $E2$ transition rate 
of $^{12}$C($\alpha,\gamma$)$^{16}$O at $E_G$, i.e., 
$E_\alpha =\frac43E_G=0.4$~MeV, the center value of $ReD_2(p)$ for 
the condition (III) becomes extremely tiny, much smaller than the errors of
$ReD_2(p)$. This will lead to a large uncertainty in an estimate of 
the $E2$ transition rate of $^{12}$C($\alpha$,$\gamma$)$^{16}$O 
with the large ANC values.

\vskip 2mm 
\noindent 
{\bf 5. Results and discussion}

In this work, we studied the elastic $\alpha$-$^{12}$C scattering 
for the $d$-wave channel at the low energy region including the first
and second $d$-wave resonant states, the $2_2^+$ and $2_3^+$ states
of $^{16}$O, in an effective Lagrangian approach. 
The phase shift $\delta_2$ is separated into three; one is
for the non-resonant part which contains the sub-threshold
$2_1^+$ state of $^{16}$O, and the other two parts are for the two resonant
$2_2^+$ and $2_3^+$ 
states of $^{16}$O. 
We include  four effective range parameters, $a_2$, $r_2$, $P_2$, $Q_2$, 
in each of the three amplitudes (thus 12 parameters in total) 
due to the modification of the counting rules discussed in the appendix
as well as in Ref.~\cite{a-prc18}, 
while we employed the nine parameters
$\{r_2, P_2,Q_2,
E_r^{(rs1)}, \Gamma_r^{(rs1)}, 
E_r^{(rs2)}, \Gamma_r^{(rs2)}, a,b\}$ 
as free parameters because $a_2$ is fixed by using the binding energy of
the $2_1^+$ state of $^{16}$O and 
two of them ($a$ and $b$ for $\tilde{A}_2^{(rs1)}$) turned out to be 
insensitive to the parameter fit. 
We also introduce a contribution from the $2_4^+$ state of $^{16}$O as a 
background from high energy using the fixed experimental energy and 
width, $E_r^{(rs3)}$ and $\Gamma_r^{(rs3)}$.
To study the issue of the scattered, large and small values of the ANC,
we introduce the three conditions (I), (II), (III) at the very
low energy region where the phase shift data are not available.
The nine parameters in the $S$ matrices with and without the $2_4^+$ state
of $^{16}$O are fitted to the phase shift data
applying one of the three conditions.  

For all the six cases of the parameter fit, the parameters are 
fitted very well to the data where the $\chi^2/N$ values are
less than one or almost one. The fitted values of the energies and widths
of the two resonant states do not depend on the choice of 
the conditions, and our fitted values of the energies and widths 
basically agree with the experimental values.
Those of the effective range parameters, $r_2$, $P_2$, $Q_2$, and the 
parameters $a$ and $b$ turned out to be sensitive to the choice
of the conditions (and $b$ is the most sensitive to the contribution 
from the $2_4^+$ state of $^{16}$O); 
they are fitted to the phase shift data at the 
non-resonance region, at the energies below the first resonant state
and between the two resonant states. By using the fitted values of the
effective range parameters, we can reproduce the both small and large
values of the ANC, $|C_b|$, depending on the choice of the conditions.
We also find the large error bars for the large ANC values. Therefore,
it is not easy to clearly pin down the value of ANC from the 
phase shift data
due to the large error bars of the ANC values and 
the $\chi^2/N$ values for the parameter fit mentioned above.  

The problem with fitting the parameters to the phase shift data for $d$-wave
channel is that though the two resonant peaks are well reproduced as seen
in Fig.~\ref{fig;del2}, it is essential to fit the parameters to the tiny
data at the low energy tail and the energy region between the two resonant
states for the study of the ANC of the $2_1^+$ state of $^{16}$O.  
Because the size of those data is so small compared 
to the distance between the binding 
energy of the sub-threshold $2_1^+$ state of $^{16}$O and the lowest energy 
of the phase shift data, $E_\alpha=2.6$~MeV, 
it is not easy to clarify how much the fitting procedures work well
when a number of parameters of a polynomial function are fitted 
to the data.
This may be seen as an ambiguity of the curves of $ReD_2(p)$ plotted in 
Fig.~\ref{fig;ReD2}.

Apart from the ambiguity mentioned above,
we reproduced the center values for the small and large ANC values 
using the conditions (I) and (III), respectively, 
in the tables \ref{table;parameters_A} and \ref{table;parameters_B}. 
Because the error bar of the large ANC turns out to be very large, 
it is not so sure that we could reproduce
the large ANC value from the phase shift data of $\alpha$-$^{12}$C 
scattering for $d$-wave channel. 
Thus, 
as discussed in the literature, 
the value of ANC of the sub-threshold $2_1^+$ state of $^{16}$O 
may need to be fixed by using other experimental data, 
such as the $\alpha$ transfer reactions,
e.g., $^{12}$C($^6$Li,$d$)$^{16}$O$^*(2_1^+)$ and 
$^{12}$C($^7$Li,$t$)$^{16}$O$^*(2_1^+)$, 
a cascade transition, $^{12}$C($\alpha$,$\gamma$)$^{16}$O$^*(2_1^+)$, and
a radiative decay of the excited state, 
$^{16}$O$^*(2_1^+)\to ^{16}$O$_{g.s.}+\gamma$,
for the present approach. 
Those data may determine the value of ANC and provide
a consistency check 
for a theoretical framework to estimate the $E2$ transition of 
$^{12}$C($\alpha$,$\gamma$)$^{16}$O at $E_G$.

\vskip 2mm \noindent
{\bf Acknowledgements}

This work was supported by the Basic Research Program through 
the National Research Foundation of Korea funded by the Ministry
of Education of Korea (NRF-2019R1F1A1040362).

\vskip 2mm 
\noindent 
{\bf Appendix}

In this appendix, we discuss a renormalization procedure 
for the effective range terms due to large and significant 
contributions from the Coulomb self-energy term, $-2\kappa H_2(p)$,
for the $d$-wave scattering.
It has been discussed for the $s$-wave scattering in Ref.~\cite{a-prc18}.

At the energy region below the resonant energies, 
where one may assume that the resonant parts of the amplitudes are negligible,
one can make a relation between the phase shift $\delta_2$ and the
non-resonant part of the amplitude as
\bea
C_\eta^2W_2(p)p\cot \delta_2 = K_2(p)-2\kappa Re H_2(p)\,,
\label{eq;Ceta2W2pcotdelta2}
\eea
where the function $K_2(p)$ is represented as 
an effective range expansion, and the Coulomb self-energy term,
$-2\kappa H_2(p)$, can be expanded in powers of $p^2/\kappa^2$ too. 
Thus, one has
\bea
-2\kappa Re H_2(p) &=& \frac{1}{24}\kappa^3p^2 
+ \frac{51}{240}\kappa p^4 
+ \frac{191}{1008}\frac{p^6}{\kappa}
- \frac{289}{10080}\frac{p^8}{\kappa^3}
+ \cdots\,.
\label{eq;-2kappaReH2}
\eea 
At the smallest energy of the experimental data, $E_\alpha = 2.6$~MeV,
where $p=\sqrt{2\mu E}=\sqrt{1.5\mu E_\alpha}$ = 104~MeV, 
one can check the series of the terms in Eq.~(\ref{eq;-2kappaReH2}) 
converge; those terms numerically become
\bea
-2\kappa Re H_2(p=104~{\rm MeV}) = 
0.022 + 0.021 + 0.003 - 0.0009 + \cdots \ \  ({\rm fm}^{-5})\,.
\label{eq;numerical_values}
\eea
While the phase shift $\delta_2$ at $E_\alpha=2.6$~MeV is 
$\delta_2=-0.0116^\circ$~\cite{tetal-prc09}, and the left-hand-side 
of Eq.~(\ref{eq;Ceta2W2pcotdelta2}) becomes
\bea
\left.
C_\eta^2 W_2(p)p\cot\delta_2 
\right|_{p=104{\rm MeV}} &=&  -0.019\ {\rm fm}^{-5}\,.
\label{eq;from_data}
\eea 
Because the first and second terms in the r.h.s 
of Eq.~(\ref{eq;numerical_values}) obtained from the Coulomb self-energy
term are in the same order of magnitude
as that calculated from the experimental phase shift data 
in Eq.~(\ref{eq;from_data}),
one needs to subtract them by including the corresponding effective range 
terms at $p^2$ and $p^4$ orders in $K_2(p)$ as counterterms.
In addition, to control sub-leading corrections 
we include the effective range term at $p^6$ order in $K_2(p)$ as well.


\vskip 3mm \noindent
\begin{thebibliography}{99}

\bibitem{f-rmp84}
W.~A. Fowler,
Rev. Mod. Phys. \textbf{56}, 149 (1984).

\bibitem{bb-npa06}
L.~R. Buchmann and C.~A. Barnes,
Nucl. Phys. A {\bf 777}, 254 (2006).

\bibitem{chk-epja15}
A. Coc, F. Hammache, J. Kiener,
Eur. Phys. J. A {\bf 51}, 34 (2015).

\bibitem{bk-ppnp16}
C.~A. Bertulani and T. Kajino,
Prog. Part. Nucl. Phys. {\bf 89}, 56 (2016).

\bibitem{detal-rmp17}
R.~J. deBoer {\it et al.},
Rev. Mod. Phys. {\bf 89}, 035007 (2017),
and references therein.

\bibitem{hkvk-rmp20}
H.-W. Hammer, S. Konig, and U. van Kolck,
Rev. Mod. Phys. \textbf{92}, 25004 (2020).

\bibitem{a-epja21}
S.-I. Ando, 
Eur. Phys. J. A \textbf{57}, 17 (2021).


\bibitem{a-prc18}
S.-I. Ando,
Phys. Rev. C \textbf{97}, 014604 (2018).

\bibitem{a-epja16}
S.-I. Ando,
Eur. Phys. J. A \textbf{52}, 130 (2016).

\bibitem{a-jkps18}
S.-I. Ando,
J. Korean Phys. Soci. \textbf{73}, 1452 (2018).

\bibitem{a-prc20}
S.-I. Ando,
Phys. Rev. C \textbf{102}, 034611 (2020).

\bibitem{a-prc19}
S.-I. Ando,
Phys. Rev. C \textbf{100}, 015807 (2019).

\bibitem{a-fbs21}
S.-I. Ando,
Few-Body Syst. \textbf{62}, 55 (2021).

\bibitem{klh-jpg13}
S. Konig, D. Lee, and H.-W. Hammer,
J. Phys. G: Nucl. Part. Phys. \textbf{40}, 045106 (2013). 

\bibitem{oin-prc16}
Yu.~V. Orlov, B.~F. Irgaziev, and L.~I. Nikitina,
Phys. Rev. C \textbf{93}, 014612 (2016).

\bibitem{bgkv-prl99}
C.~R. Brune, W.~H. Geist, R.~W. Kavanagh, and K.~D. Veal,
Phys. Rev. Lett. \textbf{83}, 4025 (1999).

\bibitem{betal-npa07}
A. Belhout et al.,
Nucl. Phys. A \textbf{793}, 178 (2007).

\bibitem{ab-plb09}
S. Adhikari and C. Basu,
Phys. Lett. B \textbf{682}, 216 (2009).

\bibitem{aetal-prl15}
M.~L. Avila et al.,
Phys. Rev. Lett. \textbf{114}, 071101 (2015).

\bibitem{s-prc04}
J.-M. Sparenberg,
Phys. Rev. C \textbf{69}, 034601 (2004).

\bibitem{dd-prc08}
M. Dufour and P. Descouvemont,
Phys. Rev. C \textbf{78}, 015808 (2008).

\bibitem{tetal-prc09}
P. Tischhauser \textit{et al.},
Phys. Rev. C \textbf{79}, 055803 (2009).

\bibitem{g-prc09}
B. Gelman, 
Phys. Rev. C \textbf{80}, 034005 (2009).


\bibitem{a-epja07}
S.-I. Ando,
Eur. Phys. J. A \textbf{33}, 185 (2007).

\bibitem{b-pr49}
H.~A. Bethe,
Phys. Rev. {\bf 76}, 38 (1949).

\bibitem{k-npb97}
D.~B. Kaplan,
Nucl. Phys. B {\bf 494}, 471 (1997).

\bibitem{bs-npa01}
S.~R. Beane and M.~J. Savage,
Nucl. Phys. A {\bf 694}, 511 (2001).

\bibitem{ah-prc05}
S. Ando and C.~H. Hyun,
Phys. Rev. C {\bf 72}, 014008 (2005).

\bibitem{g-npa04}
H.~W. Grie\ss hammer,
Nucl. Phys. A \textbf{744}, 192 (2004).

\bibitem{hfvk-epja21}
J.~B. Habashi, S. Fleming, and U. van Kolck,
Eur. Phys. J. A \textbf{57}, 169 (2021).

\bibitem{hhvk-npa08}
R. Higa, H.-W. Hammer, and U. van Kolck,
Nucl. Phys. A \textbf{809}, 171 (2008).

\bibitem{ir-prc84}
Z.~R. Iwinski and L. Rosenberg,
Phys. Rev. C \textbf{29}, 349 (1984).

\bibitem{emcee}
D. Foreman-Mackey et al., 
Pub. Astro. Soc. Pac. \textbf{125}, 306 (2013). 

\bibitem{scb-jp11}
J.-M. Sparenberg, P. Capel, and D. Baye,
Jour. Phys.: Conf. Seri. \textbf{312}, 082040 (2011). 

\bibitem{twc-npa93}
D.~R. Tilley, H.~R. Weller, and C.~M. Cheves,
Nucl. Phys. A \textbf{564}, 1 (1993).

\bibitem{ya-jkps19}
H.-E. Yoon and S.-I. Ando,
J. Korean Phys. Soci. \textbf{75}, 202 (2019).

%

\end{thebibliography}
\end{document}